\documentclass[12pt]{article}
\usepackage{epsf} 
\usepackage{amsmath}
\usepackage{amssymb}
\usepackage{epsfig}
\usepackage{latexsym}
\usepackage{amsfonts}
\usepackage{graphicx}
\usepackage{cite}
\usepackage{color}
\usepackage{url}

\begin{document}
\parindent 0mm 
\setlength{\parskip}{\baselineskip} 
\thispagestyle{empty}
\pagenumbering{arabic} 
\setcounter{page}{1}
\mbox{ }
\hfill MITP/17-045
\\
\mbox{ }
\hfill HIM-2017-04

\begin{center}
{\LARGE {\bf Anomalous magnetic moment of the muon,  
\\[1ex]
a hybrid approach}}
\\
\end{center}
\vspace{.05cm}
\begin{center}
{\bf  C. A. Dominguez}$^{(a)}$, 
{\bf H. Horch}$^{(b)}$, 
{\bf B. J\"ager}$^{(c)}$, 
{\bf N. F. Nasrallah}$^{(d)}$, 
\\
{\bf K. Schilcher}$^{(a),(e),(f)}$, 
{\bf H. Spiesberger} $^{(a),(f)}$, 
{\bf H. Wittig}$^{(b),(g)}$ 
\end{center}

\begin{center}
{\it $^{(a)}$Centre for Theoretical and Mathematical Physics, 
and Department of Physics, University of
Cape Town, Rondebosch 7700, South Africa}
\\

{\it $^{(b)}$PRISMA Cluster of Excellence, 
Institut f\"{u}r Kernphysik, 
\\ 
Johannes Gutenberg-Universit\"{a}t, D-55099 Mainz, Germany}
\\

{\it $^{(c)}$
ETH Z\"urich, Institute for Theoretical Physics, 
Wolfgang-Pauli-Str.\ 27, 
\\
8093 Z\"urich, Switzerland
}
\\

{\it $^{(d)}$
Faculty of Science, Lebanese University, Tripoli, Lebanon}
\\

{\it $^{(e)}$National Institute for Theoretical Physics, 
\\ Private Bag X1, Matielnd 0760, South Africa}
\\

{\it $^{(f)}$PRISMA Cluster of Excellence, 
Institut f\"{u}r Physik, 
\\
Johannes Gutenberg-Universit\"{a}t, D-55099 Mainz, Germany}
\\

{\it $^{(g)}$Helmholtz Institute Mainz, 
\\
Johannes Gutenberg-Universit\"{a}t, D-55099 Mainz, Germany}
\\

\end{center}
\begin{center}
\footnotesize
{\it E-mail:} 
cesareo.dominguez@uct.ac.za, karl.schilcher@uni-mainz.de, 
\\
spiesber@uni-mainz.de 
\end{center}
	
\date{\today}

\begin {abstract}
A new QCD sum rule determination of the leading order hadronic vacuum
polarization contribution to the anomalous magnetic moment of the
muon, $a_{\mu}^{\rm hvp}$, is proposed. This approach combines data on
$e^{+}e^{-}$ annihilation into hadrons, perturbative QCD and lattice
QCD results for the first derivative of the electromagnetic current
correlator at zero momentum transfer, $\Pi_{\rm EM}^\prime(0)$. The
idea is based on the observation that, in the relevant kinematic
domain, the integration kernel $K(s)$, entering the formula relating
$a_{\mu}^{\rm hvp}$ to $e^{+}e^{-}$ annihilation data, behaves like
$1/s$ times a very smooth function of $s$, the squared energy. We find
an expression for $a_{\mu}$ in terms of $\Pi_{\rm EM}^\prime(0)$,
which can be calculated in lattice QCD. Using recent lattice results
we find a good approximation for $a_{\mu}^{\rm hvp}$, but the
precision is not yet sufficient to resolve the discrepancy between the
$R(s)$ data-based results and the experimentally measured value.
\end{abstract}
\newpage


\section{Introduction}

The discrepancy between the theoretical prediction of the muon
magnetic moment anomaly, $a_{\mu}$, and its experimentally measured
value constitutes one of the few remaining problems for the Standard
Model. The main uncertainty arises from hadronic contributions. The
leading order hadronic vacuum polarization contribution can be
expressed as an integral over the total hadronic $e^{+}e^{-}$
annihilation cross section $R(s)$ multiplied by a kernel $K(s)/s$
where $s$ is the square of the center-of-mass energy. The range of
integration extends from threshold, $s_{\text{thr}}$, to infinity. The
kernel $K(s)$ is strongly peaked at small $s$ so that the integral is
dominated by the $e^+e^-$ cross section in the $\pi\pi$ channel. The
available data have recently improved significantly, leading to the
value $a_{\mu}^{\text{had}} = (693.1\pm3.4) \times 10^{-10}$
\cite{Davier:2017zfy} using the phenomenological dispersion relation
analysis. This result leads to a deviation of the Standard Model
prediction from the direct measurement of $a_{\mu}$
\cite{Olive:2016xmw} by $a_{\mu}^{\rm exp} - a_{\mu}^{\rm SM} = (26.8
\pm 7.6) \times 10^{-10}$, corresponding to a $3.5\sigma$
discrepancy. Similar findings are reported by an alternative
evaluation of the dispersion integral \cite{Hagiwara:2011af}.

In addition to phenomenological determinations of $a_\mu^{\rm hvp}$ 
relying on experimental data, it is important to exploit 
fully, or partly, some theoretical estimates of the anomaly. 
They all start with a determination of the hadronic part of 
the vacuum polarization, i.e.\ the correlator of two 
electromagnetic currents, $\Pi_{\rm EM}(s)$. Most of 
the theoretical calculations are based on QCD sum rules 
(QCDSR) \cite{Eidelman:1995ny,Groote:2003kg,Davier:2010nc, 
Hagiwara:2011af} or on lattice QCD (LQCD). The calculation 
of the correlator $\Pi_{\rm EM}(Q^2)$ in the Euclidean domain 
is a major activity within the LQCD community. Its main 
purpose is the determination of $a_\mu$ by evaluating a 
convolution integral involving the subtracted correlator 
$\Pi_{\rm EM}(Q^2) - \Pi_{\rm EM}(0)$ in the Euclidean domain, 
i.e., for $Q^2<0$ \cite{deRafael:1993za,Blum:2002ii}. A large 
part of that calculation is concerned with the determination 
of the additive renormalization, $\Pi_{\rm EM}(0)$, as well 
as the slope $\Pi_{\rm EM}^\prime(0)$. The latter gives the 
leading contribution in the representation of 
$\Pi_{\rm EM}(Q^2) - \Pi_{\rm EM}(0)$ in terms of Pad\'e 
approximants \cite{DellaMorte:2011aa,Aubin:2012me, 
Golterman:2014ksa}.  

Some time ago it was shown in Refs.\ \cite{Bodenstein:2011qy, 
Dominguez:2017omw} that a very precise approximation 
of the kernel $K(s)$ by a meromorphic function allows us 
to reduce the required theoretical information to a few 
derivatives of $\Pi_{\rm EM}^\prime(s)$ at the origin. At 
the time of this proposal, numerical predictions relied on 
model-dependent estimates of these derivatives. Since then, 
though, precise results from LQCD have become available 
\cite{Aubin:2006xv,Boyle:2011hu,Feng:2011zk,Francis:2013qna, 
Burger:2013jya,Chakraborty:2014mwa,Chakraborty:2016mwy,Borsanyi:2016lpl, 
DellaMorte:2017dyu}. LQCD results for the first derivative of 
$\Pi_{\rm EM}(s)$ at $s=0$ have been used in 
\cite{Dominguez:2017omw}, updating \cite{Bodenstein:2011qy}. 
A recent systematic approach, beyond a simple Taylor expansion 
of $\Pi_{\rm EM}(s)$, makes use of a Mellin-Barnes 
representation \cite{deRafael:2014gxa}. This allows one 
to express $a_{\mu}^{\rm hvp}$ as a series over moments 
and log-enhanced moments. It has been argued that this 
provides a tool to obtain precise results from LQCD for 
Euclidean momenta. Convergence properties and results based 
on additional model assumptions have been studied recently 
in \cite{deRafael:2017gay} and \cite{Benayoun:2016krn}. 

In this note we discuss a new hybrid method which combines 
QCDSR and LQCD, and requires the use of data on $R(s)$. 
However, the latter contributes only as a small correction 
so that data errors become quite irrelevant. There is also 
a small contribution from the asymptotic region where one 
can safely use perturbative QCD (PQCD). The most important 
contribution, by far, arises from the derivative of the 
electromagnetic correlator $\Pi_{\rm EM}^{\prime}(s=0)$ 
determined by recent LQCD calculations.


\section{Hybrid QCD sum rule}

The standard expression of the hadronic contribution to 
the muon anomaly is given by
\begin{equation}
a_{\mu} 
= \frac{\alpha_{\rm EM}^{2}}{3\,\pi^{2}} \, 
\int_{s_{\rm thr}}^{\infty} \, 
\frac{ds}{s} K(s) R(s) \, , 
\label{Eq:g1}
\end{equation}
where $\alpha_{\rm EM}$ is the electromagnetic fine structure 
constant and $R(s)$ is the bare cross section for $e^+ e^- 
\rightarrow \text{hadrons}$, normalized to the cross 
section for $\mu$-pair production including final state 
radiation. The integral starts at a threshold $s_{\rm thr} 
> 0$, which is often identified with the two-pion threshold\footnote{
  The correct threshold is $s_{\text{thr}} = m_{\pi^0}^2$ 
  for the $\pi^0 \gamma$ final state. Its contribution 
  to $a_\mu^{\rm hvp}$ is very small, but it is included in our data 
  integral (see below).  
  }, $s_{\rm thr} = 4m_{\pi}^{2}$. 
In terms of the correlator $\Pi_{\rm EM}(s)$ of two 
electromagnetic currents, we have $R(s) = 3 \, \sum_{f} \, 
Q_{f}^{2} \left[6\,\pi\operatorname{Im}\Pi_{\rm EM}(s) 
\right]$ and $\Pi_{\rm EM}(s)$ is defined as
\begin{equation}
\Pi_{\rm EM}^{\mu\nu}(q^{2}) 
= 
i \int d^{4}x \, e^{iqx} 
\langle0|T\left(j^{\mu}_{\rm EM}(x) \, 
j^{\nu}_{\rm EM}(0)\right)|0\rangle 
= 
(q_{\mu}q_{\nu} - q^{2}g_{\mu\nu}) \Pi_{\rm EM}(q^{2}) 
\, ,
\end{equation}
where $j^{\mu}_{\rm EM}(x) = \sum_c \sum_{f} Q_{f} 
\bar{q}_{f,c}(x)\gamma^{\mu}q_{f,c}(x)$ is the electromagnetic 
current and $Q_{f}$ the quark charges. The integration kernel 
$K(s)$ is given by
\begin{equation}
K(s) 
= 
\int_{0}^{1} dx \, 
\frac{x^{2}(1-x)}{x^{2}+\frac{s}{m_{\mu}^{2}}(1-x)} 
\, ,
\end{equation}
where $m_{\mu} = 0.105658$ GeV is the muon mass. An analytic 
form can be found, e.g.\ in Ref.\ \cite{Jegerlehner:2009ry} 
and will be used for our numerical evaluations. The function 
$sK(s)$ is slowly varying in the range of integration. It 
increases monotonically from $2.36 \times 10^{-3}$ GeV$^{2}$ 
at $s = 4m_{\pi}^{2}$ to $m_{\mu}^{2}/3 \simeq 3.72 \times 
10^{-3}$ at $s=\infty$. This means that the kernel $K(s)/s$ 
behaves roughly like $1/s^{2}$ over the range of integration. 
Thus the kernel gives a very large weight to the low-energy 
region, in particular to the $\rho$-meson contribution.

At first, we consider in Eq.\ (\ref{Eq:g1}) only the 
light-quark contribution. Hence, the integral extends from 
$s_{\rm thr} = 4m_{\pi}^{2} \simeq 0.078$ GeV$^{2}$ to 
$s = s_{0}$ where $s_{0}$ should be in the scaling region, 
but below the heavy quark threshold, i.e.\ in the range 
3 GeV$^2 \leq s_0 \leq 9$ GeV$^{2}$. For definiteness, we 
choose $s_{0} = 4$ GeV$^{2}$. The remaining integral from 
$s_{0}$ to $\infty$ can be safely computed in PQCD; it is 
anyway small due to the rapid fall-off of $K(s)$. 

In order to use finite energy QCD sum rules (FESR) one 
needs to approximate $K(s)$ by a meromorphic function 
$K_{1}(s)$. As $K(s)/s$ behaves approximately like $1/s^{2}$, 
we can choose, for example, 
\begin{equation}
\frac{K_{1}(s)}{s} 
= 
\frac{c_{-2}}{s^{2}} + c_{0} + c_{1}s
\quad \text{for} \quad 
s_{\rm thr} \leq s \leq s_0 \, . 
\label{Eq:K_1} 
\end{equation}
Below we will derive a sum rule for $a_{\mu}^{(u,d,s)}$ 
which is independent of the choice of the coefficients $c_i$. 
To be specific, 
we determine the constants $c_{-2}$, $c_{0}$, and $c_{1}$ by 
the conditions 
\begin{equation}
\int_{s_{\rm thr}}^{s_0} 
\frac{K(s)}{s} s^n ds 
= 
\int_{s_{\rm thr}}^{s_0} 
\frac{K_{1}(s)}{s} s^n ds 
\label{Eq:fixci}
\end{equation}
for $n=0, 1, 2$. This leads to 
\begin{equation}
c_{-2} = 2.762 \times10^{-3}~{\rm GeV}^2, \quad
c_{0} = 4.136 \times10^{-4}~{\rm GeV}^{-2}, \quad 
c_{1} = -9.914 \times10^{-5}~{\rm GeV}^{-4} \, . 
\label{Eq:K_1_a} 
\end{equation}
We will use this approximation in subsequent sections 
to illustrate the relative size of different contributions 
to $a_{\mu}$. However, we will also consider other possible 
choices for the approximate kernel. 

For three flavors the integral in Eq.\ (\ref{Eq:g1}) can be 
separated into a low- and a high-energy part. In the 
low-energy part we add and subtract the approximation 
$K_1(s)$ to obtain 
\begin{align*}
a_{\mu}^{(u,d,s)} 
& = 
\frac{\alpha_{\rm EM}^{2}}{3\,\pi^{2}} \, 
\int_{4m_{\pi}^{2}}^{s_0} 
\frac{ds}{s} \; 
\left[K(s) - K_{1}(s) + K_{1}(s)\right] \; 12\pi 
\, \operatorname{Im}\Pi_{\rm EM}(s) 
\\
& + 
\frac{\alpha_{\rm EM}^{2}}{3\,\pi^{2}} \, 
\int_{s_0}^{\infty}\frac{ds}{s} \; 
K(s)R(s)
\end{align*}
Subsequently, we re-write the low-energy part in terms 
of $K_1(s)$: 
\begin{align}
& 
\int_{4m_{\pi}^{2}}^{s_{0}} 
\frac{ds}{s} K_{1}(s) \, 
\operatorname{Im}\Pi_{\rm EM}(s) 
\nonumber 
\\
& = 
- \frac{1}{2 i} 
\oint_{|s|=s_{0}} \frac{ds}{s} 
K_{1}(s) \Pi_{\rm EM}(s) 
+ \, \pi \text{Res} 
\left[\frac{1}{s} K_{1}(s) \Pi_{\rm EM}(s) \right]_{s=0} 
\nonumber 
\\
& =  
- \frac{1}{2 i} 
\oint_{|s|=s_{0}} \frac{ds}{s} \; K_{1}(s) \Pi_{\rm EM}(s) 
+ \pi c_{-2}\Pi^{\prime}_{\rm EM}(0)  
\, ,
\label{Eq:g12} 
\end{align}
where the integral around the circle of radius $s_{0}$ can 
be computed using PQCD. 

We have thus four contributions to the anomalous magnetic 
moment of the muon: 
\begin{equation}
a_\mu^{(u,d,s)} 
= a_\mu^{(1)} + a_\mu^{(2)} + a_\mu^{(3)} + a_\mu^{(4)}
\label{Eq:hsumrule}
\end{equation}
where 
\begin{eqnarray}
a_\mu^{(1)}
&=& 
\frac{\alpha_{\rm EM}^{2}}{3\,\pi^{2}} \,
6\pi i 
\oint_{|s|=s_{0}} \frac{ds}{s} \; K_{1}(s) \Pi_{\rm EM}(s) 
\, , 
\label{Eq:amu1} 
\\
a_\mu^{(2)} 
&=& 
\frac{\alpha_{\rm EM}^{2}}{3\,\pi^{2}} \,
12\pi^2 c_{-2}\Pi^{\prime}_{\rm EM}(0)
\, , 
\label{Eq:amu2} 
\\
a_\mu^{(3)} 
&=& 
\frac{\alpha_{\rm EM}^{2}}{3\,\pi^{2}} \,
\int_{4m_{\pi}^{2}}^{s_0} \, 
\frac{ds}{s} \; [K(s)-K_{1}(s)] \; R(s) 
\, , 
\label{Eq:amu3} 
\\
a_\mu^{(4)} 
&=& 
\frac{\alpha_{\rm EM}^{2}}{3\,\pi^{2}} \,
\int_{s_0}^{\infty} \, 
\frac{ds}{s} \; K(s) \; R(s) 
\, . 
\label{Eq:amu4} 
\end{eqnarray}
Only the third part, $a_\mu^{(3)}$ will be calculated 
using experimental data, with the others obtained entirely 
from theory. It turns out that the data-dependent contribution 
is only a small correction. Experimental uncertainties are 
consequently considerably suppressed. It is important to 
emphasize that the sum rule Eq.\ (\ref{Eq:hsumrule}) is exact. 
The precision of the approximation $K_1(s)$ is not relevant 
for the total value of $a_\mu^{(u,d,s)}$. This is because 
what is included is the difference $K(s) - K_1(s)$. A good 
approximation to the kernel will, however, reduce the impact 
of data uncertainties entering $a_\mu^{(3)}$. 

The first derivative of $\Pi_{\rm EM}$ at $s=0$ enters in 
Eq.\ (\ref{Eq:amu2}) because the approximate kernel $K_1(s)$ 
contains a term proportional to $1/s$. In principle it is 
possible to include higher inverse powers of $s$, as e.g.\ 
in Ref.\ \cite{Bodenstein:2011qy}, where a very good 
approximation of the kernel $K(s)$ was found using 
powers $s^{-n}$ up to $n=3$. There, the data integral, 
$a_\mu^{(3)}$ turned out to be completely negligible, at 
the level below one permille. Instead, higher 
derivatives of $\Pi_{\rm EM}$ would contribute: 
\begin{equation}
\frac{K_1(s)}{s} 
= \sum_{n\geq 2} \frac{c_{-n}}{s^n} + c_0 + c_1 s
\quad \longrightarrow \quad 
a_\mu^{(2)} 
= 
\frac{\alpha_{\rm EM}^{2}}{3\,\pi^{2}} \, 12\pi^2 
\sum_{n\geq 2} 
\frac{c_{-n}}{(n-1)!} 
\left.
\left(\frac{d}{ds}\right)^{n-1}\Pi_{\rm EM}(s)
\right|_{s=0}
\, .
\end{equation}
With such a modification of our approach one would reduce 
the contributions which can be calculated in PQCD and from 
data in favor of non-perturbative input from LQCD. Sufficiently 
precise results for the higher derivatives from LQCD are, 
however, not yet available and we will not study this 
possibility further. Next, one has to add the heavy quark contributions. These will be obtained from PQCD as described 
below. In subsequent sections we will discuss each term in 
Eq.\ (\ref{Eq:hsumrule}).


\section{Calculation of the low-energy PQCD part: 
{\boldmath $a_\mu^{(1)}$}}

The integral in Eq.\ (\ref{Eq:amu1}) involving $\Pi_{\text{EM}}$ 
contains a dominant purely perturbative term, plus corrections 
due to condensates and duality violations. In PQCD, the 
current-current correlator is usually defined for the vector 
current of a single quark type, so that for three light 
flavors, using $3\sum Q_{f}^{2} = 2$, one has $\Pi_{\text{EM}} = 
(2/3) \Pi^{\text{PQCD}}$. We calculate the integral containing 
$\Pi^{\text{PQCD}}$ using fixed-order perturbation theory in 
the light-quark sector at five-loop level. To achieve this we 
make use of moments defined by 
\begin{equation}
M_{N}(s_{0}) 
= 
4\pi^{2} \int\limits_{0}^{s_{0}} \frac{ds}{s_{0}} 
\left[\frac{s}{s_{0}}\right]^{N} 
\frac{1}{\pi}\operatorname{Im}\Pi^{\text{PQCD}}(s) 
= 
\frac{-1}{2\pi i}\oint_{|s|=s_{0}} \frac{ds}{s_{0}} 
\left[\frac{s}{s_{0}}\right]^{N} 
4\pi^{2}\Pi^{\text{PQCD}}(s) 
\, ,
\end{equation}
with \cite{Baikov:2008jh}
\begin{eqnarray}
4\pi^{2}\Pi^{\text{PQCD}}(s) 
& = & 
- L - La - a^{2} \left(Lk_{2} - \frac{1}{2}L^{2}\beta_{0}\right) 
\label{Eq:Pipqcd}
\\
& & - a^{3}\left(Lk_{3} + \frac{1}{3}L^{3}\beta_{0}^{2} + 
L^{2}\left(-\frac{1}{2}\beta_{1} - \beta_{0}k_{2}\right) \right) 
\nonumber
\\
& & - a^{4}\left(Lk_{4} - \frac{1}{4}L^{4}\beta_{0}^{3} 
+ \frac{1}{6}L^{3}\beta_{0}\left(5\beta_{1} + 6\beta_{0}k_{2}\right)  
\right.
\nonumber
\\
& & \left.
+ L^{2}\left(- \frac{1}{2}\beta_{2} - \frac{3}{2}\beta_{0}k_{3} 
- \beta_{1}k_{2}\right)\right)
\nonumber
\end{eqnarray}
where $L = \ln(-s/\mu^2)$ and $a = \alpha_s({\mu^2})/\pi$ is 
the running strong coupling at a renormalization scale  
$\mu^2$. The constants are $k_{1} = 1$, $k_{2} = 1.6398$, 
$k_{3} = 6.3711$, $k_{4} = 49.076$, $\beta_{0} = \frac{9}{4}$, 
$\beta_{1} = 4$, $\beta_{2} = 10.060$, $\beta_{3} = 47.228$. 
The result for the low-energy PQCD contribution to the 
anomaly is 
\begin{equation}
a_\mu^{(1)}
= 
\frac{\alpha_{\rm EM}^{2}}{3\,\pi^{2}} \, 
2 s_{0} \left[\frac{c_{-2}}{s_{0}^{2}}M_{-2}(s_0) 
+ c_{0}M_{0}(s_0) + c_{1}s_{0}M_{1}(s_0) \right]  
\, . 
\end{equation}
Explicit expressions for the moments can be found in Ref.\ 
\cite{Penarrocha:2001ig} (see also \cite{Beneke:2008ad}). 
We use $\alpha_s(M_Z)=0.1181 \pm 0.0011$ \cite{Olive:2016xmw}, 
corresponding to $\alpha_{s}(m_{\tau}^{2}) = 0.321 \pm 0.009$.  
With $\mu^2 = s_0 = 4$ GeV$^2$ we obtain 
\begin{equation*}
M_{-2} = -1.0774 \pm 0.0010, \quad 
M_{0} = 1.1433 \pm 0.0052, \quad 
M_{1} = 0.5559 \pm 0.0013 \, .
\end{equation*}
With the values for $c_{-2}$, $c_{0}$ and $c_{1}$ from 
Eq.\ (\ref{Eq:K_1_a})) the low-energy PQCD part of $a_\mu^{(1)}$ 
becomes 
\begin{equation}
a_{\mu}^{(1)} 
= 
(9.56 \pm 0.21) \times 10^{-10} 
\, .
\end{equation}
Alternatively, using contour-improved perturbation theory 
\cite{LeDiberder:1992jjr,Pivovarov:1991rh}, where the 
running (i.e.\ $s$-dependent) strong coupling constant 
$\alpha_s(\mu^2 = s)$ is used inside the integrals 
for the moments, we obtain $M_{-2} = -1.0794$, $M_{0} = 
1.1372$, $M_{1} = 0.5509$ and $a_{\mu}^{(1)} = 9.44 
\times 10^{-10}$. 

This contribution to $a_{\mu}$ is small and its uncertainty 
due to scale variations or the uncertainty from $\alpha_s$ is 
negligible for the total. Finally, we consider additional 
uncertainties due to higher-dimensional operators, e.g.\ 
the gluon condensate, and due to duality violations. The 
contribution of the gluon condensate is given by 
\begin{equation}
\Pi_V^{\text{GG}}(s) 
= \frac{1}{s^{2}} \sum_{f=u,d,s} Q_{f}^{2} 
\frac{1}{12} \left(1 + \frac{7}{6}a\right) 
\left\langle \frac{\alpha_{s}}{\pi} G^2 \right\rangle 
\, , 
\label{Eq:piggv}
\end{equation}
leading to 
\begin{equation}
a_{\mu}^{(1)\text{GG}} 
= 
\frac{\alpha_{\rm EM}^{2}}{3\,\pi^{2}} 
\frac{-1}{2\pi i}\oint_{|s|=s_{0}} ds\, c_{1} 
8\pi^{2}\Pi^{\text{GG}}_V(s)
\, .
\end{equation} 
The numerical value of the gluon condensate is known with 
large uncertainties. For illustration we use a conservative 
estimate based on Ref.\ \cite{Dominguez:2014fua}, 
$\left\langle \frac{\alpha_{s}}{\pi}G^{2} \right\rangle 
= 0.015 \pm 0.015$\, GeV$^{4}$, which covers most of the 
phenomenological determinations. This leads to
\begin{equation}
a_{\mu}^{(1)\, \text{GG}} 
= (0.12 \pm 0.12) \times 10^{-10} 
\end{equation}
which is two orders of magnitude smaller than $a_{\mu}^{(1)}$ 
from PQCD. If $K_1(s)$ does not contain a linear term, i.e.\ 
if $c_1=0$ in Eq.\ \ref{Eq:K_1}, there is still a contribution 
from the higher-order term in Eq.\ (\ref{Eq:piggv}) due to the 
logarithmic $s$-dependence of $\alpha_s$. This contribution 
to $a_{\mu}^{(1)}$ will be even smaller since it is suppressed 
by an extra power of $\alpha_s$. We have also checked that 
strange-quark mass effects are tiny ($a_{\mu}^{(1)\, 
\text{strange}} = 0.05 \times 10^{-10}$ for $m_s = 0.1$ GeV) 
and hence can safely be neglected. Finally, also duality 
violations are expected to be very small due to their expected 
exponential fall-off with increasing energy. Indeed, using 
the model described in Ref.\ \cite{Pich:2016bdg} we can estimate 
their contribution as $a_{\mu}^{(1) \text{DV}} \simeq 0.06 \times 
10^{-10}$.


\section{Contribution of the pole residue part: 
{\boldmath $a_\mu^{(2)}$}}

In the present hybrid approach, the main contribution to 
$a_{\mu}$ is due to the residue term involving the first 
derivative of the electromagnetic correlator at the origin
\begin{equation}
a_{\mu}^{(2)} 
= \frac{\alpha_{\rm EM}^{2}}{3\,\pi^{2}} \, 12\pi^2 c_{-2} 
\Pi^{\prime}_{\rm EM}(0) 
\, .
\end{equation}
This quantity, dominated by non-perturbative physics, can be
determined using recent results obtained in LQCD
\cite{Chakraborty:2016mwy,Borsanyi:2016lpl,DellaMorte:2017dyu}. These
calculations are afflicted with a number of systematic uncertainties
(such as discretization artefacts, finite-volume effects or
isospin-breaking contributions) which must be brought under
sufficient control in order to achieve a competitive determination
of $a_\mu^{\rm hvp}$. The slope at the origin, $\Pi_{\rm EM}^\prime(0)$, can be
obtained by performing a fit to the numerical data for $\Pi_{\rm
  EM}(Q^2)$.  Alternatively, its value is accessible by computing the
second time moment of the spatially summed vector correlator
\cite{Chakraborty:2014mwa}.

In Table~\ref{tab:slope} we show a compilation of results for
$\Pi_{\rm EM}^\prime(0)$ from LQCD. While the three calculations
produce results in the same ballpark, they differ in several technical
aspects. In Refs.\ \cite{Chakraborty:2016mwy} and
\cite{Borsanyi:2016lpl} staggered quarks are employed to discretize
the QCD action, and the slope is determined by evaluating the second
time moment of the vector correlator. By contrast, Mainz/CLS
\cite{DellaMorte:2017dyu} use Wilson quarks and perform a low-order
Pad\'e fit to determine the slope. Still, the results listed in
Table~\ref{tab:slope} differ significantly, given the quoted
errors. The most likely explanation is that systematic uncertainties
are not sufficiently controlled in some (or all) of these
calculations. Theoretical predictions for the hadronic vacuum
polarization contribution to the muon $g-2$ obtained on the basis of
these results require a critical assessment.

\begin{table}[t]
\begin{center}
\begin{tabular}{c l l}
\hline\hline
$\Pi_{\rm EM}^\prime(0)$ & Collab. & $a$\,[fm] \\
\hline
0.0883(59) & Mainz/CLS \cite{DellaMorte:2017dyu}  & C.L. \\
0.0959(30) & BMW       \cite{Borsanyi:2016lpl}    & C.L. \\
0.0889(16) & HPQCD     \cite{Chakraborty:2016mwy} & 0.15 \\
0.0892(14) & HPQCD     \cite{Chakraborty:2016mwy} & 0.12 \\
\hline\hline
\end{tabular}
\caption{
\label{tab:slope} 
{\small 
Recent results for $\Pi_{\rm EM}^\prime(0)$ computed in LQCD 
at the physical pion mass. Results labelled by ``C.L.'' have 
been extrapolated to the continuum limit of vanishing lattice 
spacing, $a=0$. In this case the uncertainty on 
$\Pi_{\rm EM}^\prime(0)$ includes a contribution from the 
continuum limit extrapolation. The result of Ref.\ 
\cite{Borsanyi:2016lpl} includes a negative contribution 
from quark-disconnected contractions, and the (small) contribution 
from charm quarks contained in that result has been subtracted.
}}
\end{center}
\end{table}

Using the result from Ref.\ \cite{DellaMorte:2017dyu} to 
estimate the slope parameter we find $\Pi^{\prime}_{\rm EM}(0) 
= (0.0883 \pm 0.0059)~\mbox{GeV}^{-2}$, which implies that 
the residue part of the anomaly is evaluated as $a_{\mu}^{(2)} 
= (519.7 \pm 34.6) \times 10^{-10}$. Thus, the residue term 
constitutes the largest contribution to $a_{\mu}^{(2)}$ by far. 
However, it also makes the largest contribution to the error, 
and this remains true if $\Pi_{\rm EM}^\prime(0)$ is replaced by the
estimates from Refs.\ \cite{Chakraborty:2016mwy} or \cite{Borsanyi:2016lpl}.


\section{Data integral contribution: {\boldmath $a_\mu^{(3)}$}}

In this hybrid approach the integral over the data is 
\begin{equation}
a_{\mu}^{(3)} 
= 
\frac{\alpha_{\rm EM}^{2}}{3\,\pi^{2}} \, 
\int_{4m_{\pi}^{2}}^{s_0} \, \frac{ds}{s} \; 
[K(s)-K_{1}(s)] \; R(s) 
\, .
\label{Eq:amu3i}
\end{equation}
For $R(s)$ we use a previous compilation of data from  
\cite{Bodenstein:2011hm}, which was already used in 
\cite{Bodenstein:2013flq}. The result for $s_0 = 4$ GeV$^2$ 
is 
\begin{equation}
a_{\mu}^{(3)} = (55.5 \pm 0.6) \times 10^{-10}
\, .
\end{equation}
Hence, in this hybrid approach, the data contribution to the 
anomaly is small, i.e.\ around $8\, \%$ of the total hadronic 
contribution. Data errors are reduced correspondingly, and 
details of the handling of data and their uncertainties is 
presently irrelevant.


\section{High-energy asymptotic contribution: 
{\boldmath $a_\mu^{(4)}$}}

A numerical integration of the high-energy contribution, 
Eq.\ (\ref{Eq:amu4}), using Eq.\ (\ref{Eq:Pipqcd}), 
is straightforward. We set $\mu^2 = s$ and use the running 
$\alpha_s$ from {\tt RunDec} \cite{Chetyrkin:2000yt} with 
$\alpha_s(M_Z) = 0.1181$. We find 
\begin{equation}
a_\mu^{(4)} = 35.37 \times 10^{-10}
\end{equation} 
with a negligible error of $\pm 0.05 \times 10^{-10}$ due to 
the uncertainty in $\alpha_s$. Varying the renormalization 
scale $\mu$ in Eq.\ (\ref{Eq:Pipqcd}) up and down by a factor 
of two, increases $a_\mu^{(4)}$ by $+0.61$ and $+0.44$, 
respectively. 

We have checked that at high energies the kernel $K(s)$ 
can be safely approximated by its asymptotic form $K(s) 
\simeq \frac{m_{\mu}^{2}}{3s}$, with a precision of better 
than 2\,\%. Hence, $a_\mu^{(4)}$ can be calculated 
analytically and expressed in terms of a PQCD moment: 
\begin{eqnarray}
a_\mu^{(4)} 
& = & 
\left(\frac{\alpha_{\rm EM}m_{\mu}}{3\,\pi}\right)^{2} \, 2 
\int_{s_0}^{\infty} \frac{ds}{s^{2}} \; 
4\pi^{2}\frac{1}{\pi} \operatorname{Im} \Pi^{\text{PQCD}}(s) 
\nonumber \\
& = & 
- \left(\frac{\alpha_{\rm EM}m_{\mu}}{3\,\pi}\right)^{2} 
\frac{2}{s_{0}}  
M_{-2}(s_0)
\\
& = & 
36.05 \times 10^{-10} \, . 
\nonumber 
\end{eqnarray}

The use of PQCD at squared energies above $s_0 = 4$ GeV$^2$ 
is well justified. For instance, the excellent agreement of 
$R(s)$ with PQCD is supported by the recent KEDR data 
\cite{Anashin:2015woa,Anashin:2016hmv} in the range $\sqrt{s} 
= 1.84 - 3.05$ GeV. Moreover, since $a_\mu^{(4)}$ is 
only a small contribution to the total $a_\mu^{\rm hvp}$, one can 
expect corrections due to condensates and duality 
violations to be completely negligible. In particular, the 
imaginary part of $\Pi_V^{\text{GG}}(s)$, see Eq.\ 
(\ref{Eq:piggv}), is generated only by the logarithmic 
scale dependence of $\alpha_s$, hence it is suppressed by 
two powers of the strong coupling.


\section{Combined results and impact of kernel variations} 

For $s_0 = 4$ GeV$^2$ and using the estimate for $\Pi_{\rm
  EM}^\prime(0)$ derived from the LQCD results in
\cite{DellaMorte:2017dyu} we find $a_\mu^{(u,d,s)} = 
(620.1 \pm 34.6) \times 10^{-10}$, while with 
$\Pi_{\rm EM}^\prime(0)$ from \cite{Borsanyi:2016lpl} we have 
$a_\mu^{(u,d,s)} = (664.9 \pm 17.7) \times 10^{-10}$. These 
results fall in the right ballpark, but are obviously not yet 
competitive with other determinations. The slight disagreement 
between the two values shows that a careful assessment of
the error contributions to the LQCD results for $\Pi_{\rm
  EM}^\prime(0)$ is necessary. At any rate, the error in the final
result is completely dominated by the uncertainty in the LQCD
determination of the slope $\Pi_{\rm EM}^\prime(0)$.

The final result for $a_\mu^{(u,d,s)}$ should not depend on the
specific choice of the approximate kernel $K_1(s)$, if all
contributions could be calculated with the same exact current
correlator. However, differences can occur to the extent that the
different pieces of $\Pi_{\rm EM}$ are affected by different errors
and uncertainties. In fact, (i) the PQCD parts are not exact because
higher-order terms are missing, and the known parts depend on the
choice of the renormalization scale. (ii) The LQCD contribution has
its own specific statistical and systematic uncertainties associated
with the lattice approach. Furthermore, strong and electromagnetic
sources of isospin breaking are unaccounted for in Refs.\ 
\cite{Borsanyi:2016lpl,DellaMorte:2017dyu}. Recent calculations
\cite{Boyle:2017gzv,Giusti:2017jof} have provided indications that
isospin-breaking effects in $a_\mu^{\rm hvp}$ are of the order of $-1$\%.
(iii) PQCD does not take into account QED corrections. 
(iv) The data are obviously affected 
by experimental uncertainties. It is therefore interesting to 
study variations in the approximate kernel $K_1(s)$. Resulting 
shifts of $a_\mu^{(u,d,s)}$ can then be taken as an indication 
of the presence of unknown systematic errors in the various 
ingredients of our approach. 

\begin{table}[t]
\begin{center}
{\footnotesize
\begin{tabular}{|r|r|c|c|r||c|}
\hline
\rule[-3mm]{0mm}{9mm}
Case
& $a_\mu^{(1)}$
& $a_\mu^{(2)}$ (\cite{Borsanyi:2016lpl}, \cite{DellaMorte:2017dyu})
& $a_\mu^{(3)}$
& $a_\mu^{(4)}$
& $a_\mu^{(u,d,s)}$ (\cite{Borsanyi:2016lpl}, \cite{DellaMorte:2017dyu})
\\[5pt]
\hline
\rule[-3mm]{-4pt}{9mm}
$0$ 
& $9.56$ 
& ($564.5 \pm 17.7$, ~ $519.7 \pm 34.6$) 
& $55.5 \pm 0.6$
& $35.37$ 
& ($664.9 \pm 17.7$, ~ $620.1 \pm 34.6$) 
\\[5pt]
$1$ 
& $36.48$ 
& ($555.7 \pm 17.4$, ~ $511.5 \pm 34.2$) 
& $56.6 \pm 0.6$
& $15.64$ 
& ($664.4 \pm 17.4$, ~ $620.2 \pm 34.2$) 
\\[5pt]
$2$ 
& $-47.13$ 
& ($482.3 \pm 15.1$, ~ $443.9 \pm 29.6$) 
& $195.0 \pm 2.0$
& $35.37$ 
& ($665.5 \pm 15.2$, ~ $627.1 \pm 29.7$) 
\\[5pt]
$3$ 
& $-57.26$ 
& ($586.0 \pm 18.4$, ~ $539.3 \pm 36.0$) 
& $96.9 \pm 1.0$
& $35.37$ 
& ($661.0 \pm 18.4$, ~ $614.3 \pm 36.0$) 
\\[5pt]
$4$ 
& $-31.30$ 
& ($660.1 \pm 20.6$, ~ $607.8 \pm 40.7$) 
& $0$
& $35.37$ 
& ($664.2 \pm 20.6$, ~ $611.9 \pm 40.7$) 
\rule[-4mm]{-4pt}{7mm}
\\
\hline
\end{tabular}
}
\end{center}
\caption{
\label{Tab:amui}
\small{
The light-quark hadronic contribution to the muon anomalous 
magnetic moment and its split-up as defined in the text, in 
units of $10^{-10}$. In the line labeled 'Case 0' we collect 
the numerical results discussed in previous sections while 
cases 1 -- 4 are modifications of the kernel $K_1(s)$ 
described in the text. For the residue part we show two results 
based on input from LQCD of Refs.\ \cite{DellaMorte:2017dyu} and 
\cite{Borsanyi:2016lpl}. Uncertainties for the perturbative 
parts are not shown since they are sub-dominant. 
}}
\end{table}

To this end we consider four modifications of $K_1(s)$, 
to wit: 
\begin{itemize}
\item[1.]
Choosing the larger value for $s_0 = 9$ GeV$^2$ (instead 
of $s_0 = 4$ GeV$^2$) will extend the range of energies 
for which the data integral $a_\mu^{(3)}$ has to be 
evaluated. The PQCD contribution is expected to be even 
more reliable. The fit is given by Eq.\ (\ref{Eq:K_1}) 
with 
\begin{equation}
c_{-2} = 2.7195 \times 10^{-3}~{\rm GeV}^{2} , \quad 
c_{0} = 2.9618 \times 10^{-4}~{\rm GeV}^{-2} , \quad 
c_{1} = -3.6775 \times 10^{-5}~{\rm GeV}^{-4} 
\, .
\end{equation}
Both the data integral and the residue term change only 
little: 
$a_\mu^{(1)} = (36.48 \pm 0.19) \times 10^{-10}$, 
$a_\mu^{(2)} = (511.5 \pm 34.2) \times 10^{-10}$ 
(we use the LQCD result of \cite{DellaMorte:2017dyu} for the 
four cases with modified $K_1(s)$),  
$a_\mu^{(3)} = (56.6 \pm 0.6) \times 10^{-10}$,  
$a_\mu^{(4)} = (15.64 \pm 0.19) \times 10^{-10}$, and  
the total for $s_0 = 9$ GeV$^2$ is
\begin{equation}
a_\mu^{(u,d,s)} = (620.2 \pm 34.3) \times 10^{-10}
\, .
\end{equation}
\item[2.]
The ansatz 
\begin{equation}
K_1(s) = \frac{c_{-2}}{s} \left(1 - \frac{s^2}{s_0^2}\right)
\label{eq:K_1var}
\end{equation}
provides pinching at $s = s_0$ which we choose again as
$s_0 = 4$ GeV$^2$. We choose $c_{-2} = 2.36 \times 
10^{-3}~{\rm GeV}^{2}$ such that $K_1(s)$ and $K(s)$ agree 
at threshold and find 
$a_\mu^{(1)} = (-47.13 \pm 0.14) \times 10^{-10}$, 
$a_\mu^{(2)} = (443.9 \pm 29.6) \times 10^{-10}$, 
$a_\mu^{(3)} = (195.0 \pm 2.0) \times 10^{-10}$, 
$a_\mu^{(4)} = (35.36 \pm 0.61) \times 10^{-10}$, 
and for the total: 
\begin{equation}
a_\mu^{(u,d,s)} = (627.1 \pm 31.2) \times 10^{-10}. 
\end{equation}
\item[3.] 
The same form, Eq.\ (\ref{eq:K_1var}), with pinching at 
$s = s_0 = 4$ GeV$^2$, but restricting the energy range 
to $s \geq 0.2$ GeV$^2$ leads to a larger value 
for the parameter $c_{-2} = 2.87 \times 10^{-3}~{\rm GeV}^{2}$. 
This results in $a_\mu^{(1)} = (-57.26 \pm 0.17) 
\times 10^{-10}$ for fixed-order perturbation theory (or 
$a_\mu^{(1)} = (-57.19 \pm 0.17) \times 10^{-10}$ for CIPT), 
$a_\mu^{(2)} = (539.3 \pm 36.0) \times 10^{-10}$, 
$a_\mu^{(3)} = (96.9 \pm 1.0) \times 10^{-10}$, 
$a_\mu^{(4)} = (35.36 \pm 0.61) \times 10^{-10}$, 
and for the total: 
\begin{equation}
a_\mu^{(u,d,s)} = (614.3 \pm 36.3) \times 10^{-10}. 
\end{equation} 
\item[4.] 
Finally, an extreme option could be to choose $c_0 = c_1 = 0$ 
and determine the coefficient $c_{-2}$ as in Eq.\ (\ref{Eq:fixci}), 
but including $R(s)$ in the integral. This renders the data 
integral $a_\mu^{(3)}$ equal to zero by definition and the total 
$a_\mu^{(u,d,s)}$ as well as its error is dominated by the 
LQCD part. We find $c_{-2} = 3.23 \times 10^{-3}~{\rm GeV}^{2}$ 
and $a_\mu^{(u,d,s)} = (611.9 \pm 40.7) \times 10^{-10}$. 
\end{itemize}

The changes of individual contributions to $a_\mu^{(u,d,s)}$ 
(see Table \ref{Tab:amui}) compensate each other in the total 
and the final result varies little within the given 
uncertainties. In Table \ref{Tab:amui} we also show 
results where the LQCD input is taken from 
\cite{Borsanyi:2016lpl}. 

The total error is dominated by the LQCD calculation, while 
uncertainties arising from variations in the treatment of 
the kernel function and the data integral are clearly 
sub-dominant. Case 2 based on a kernel with pinching is 
preferred over the other scenarios since it minimizes the 
LQCD contribution, as well as the uncertainty. With a 
future improved determination of $\Pi^{\prime}_{\rm EM}(0)$ 
on can use  
\begin{equation}
a_\mu^{(u,d,s)} = \left(
183.2 \pm 2.1 + 5027 \, \Pi^{\prime}_{\rm EM}(0) \, {\rm GeV}^2  
\right) \times 10^{-10} 
\end{equation}
to calculate directly the light-quark contribution to 
the muon anomalous magnetic moment. 

At this point heavy-quark contributions, which can be calculated from
PQCD, have to be added. There is no need to invoke model assumptions
or to use LQCD results. Charm and bottom contributions can be
described in the same way as above. As shown in
Ref.\ \cite{Bodenstein:2011qy} excellent approximations for the kernel
$K(s)$ can be found in the respective energy ranges for the charm and
bottom sector. Therefore we take numerical results from
\cite{Bodenstein:2011qy}, for charm quarks $a_\mu^{(c)} = (14.4 \pm
0.1) \times 10^{-10}$ and for bottom quarks $a_\mu^{(b)} = (0.29 \pm
0.01) \times 10^{-10}$. The sum, $a_\mu^{(c+b)} = (14.7 \pm 0.1)
\times 10^{-10}$, has to be added to the total light-quark
contribution shown in the last columns of Table \ref{Tab:amui}. We
note that the above value of $a_\mu^{(c)}$ agrees very well with the
recent determination in LQCD \cite{DellaMorte:2017dyu},
i.e. $a_\mu^{(c)} = (14.3 \pm 0.2\pm0.1) \times 10^{-10}$.


\section{Summary}

In this paper we discussed a new QCD sum rule determination 
of the leading order hadronic vacuum polarization contribution to the anomalous 
magnetic moment of the muon, $a_\mu$. This determination 
combines theoretical input from both perturbative as well 
as lattice QCD. The splitting into different contributions 
depends on an approximation to the original kernel function, 
$K(s)$, entering the expression of the anomaly, $a_\mu^{\rm hvp}$,  
Eq.\ (\ref{Eq:g1}), involving the electromagnetic current 
correlator. This approximate kernel, $K_1(s)$, is chosen so 
as to suppress the contribution of the experimental data, 
and their uncertainties. It is also designed to allow for 
the main contribution to the anomaly to be determined by 
the first derivative of the electromagnetic current correlator 
at zero-momentum, $\Pi^{\prime}_{\rm EM}(0)$. The latter can be 
obtained from LQCD, so that the final result becomes 
essentially a QCD prediction. Several options for the 
approximate kernel $K_1(s)$ have been considered, leading 
to practically the same results for the anomaly. Regarding 
the uncertainties involved in this approach, they are 
essentially of two types. The first, and by far the largest, 
is due to the LQCD input for $\Pi^{\prime}_{\rm EM}(0)$. Secondly, 
the $e^+ e^-$ data errors, the PQCD term, and the heavy-quark 
contribution to the sum rule involve tiny uncertainties which are
comparable, if not smaller than those from the best 
data-driven determinations of $a_\mu^{\rm hvp}$. It should be mentioned 
that uncertainty sources in the PQCD part such as e.g.\ 
isospin breaking of order $(m_{u}-m_{d})^{2}/s_{0}$, the 
singlet contribution of order $m_{s}^{2}$, and potential 
duality violations are all negligible. 

Currently, LQCD uncertainties on $\Pi^{\prime}_{\rm EM}(0)$ are 
large, so that predictions from this method do not yet 
compete in accuracy with results using $e^+ e^-$ data. 
Continuous improvement in the accuracy of LQCD results for 
$\Pi^{\prime}_{\rm EM}(0)$ are likely to render this method 
competitive, thus contributing to settle the issue of 
whether $a_\mu$ opens a window beyond the Standard Model.


\section*{Acknowledgments} 
Supported in part by the National Research Foundation, NRF, 
(South Africa), the National Institute for Theoretical 
Physics, NITheP, (South Africa), and by the DFG through SFB\,1044.
HH, BJ and HW thank their colleagues within the Mainz/CLS effort for
the collaboration on Ref.\ \cite{DellaMorte:2017dyu}.




\begin{thebibliography}{99} 

\bibitem{Davier:2017zfy}
  M.~Davier, A.~Hoecker, B.~Malaescu and Z.~Zhang,
  arXiv:1706.09436 [hep-ph].

\bibitem{Olive:2016xmw}
  C.~Patrignani {\it et al.} [Particle Data Group],
  Chin.\ Phys.\ C {\bf 40} (2016) no.10,  100001.

\bibitem{Hagiwara:2011af}
  K.~Hagiwara, R.~Liao, A.~D.~Martin, D.~Nomura and T.~Teubner,
  J.\ Phys.\ G {\bf 38} (2011) 085003
  doi:10.1088/0954-3899/38/8/085003
  [arXiv:1105.3149 [hep-ph]].

\bibitem{Eidelman:1995ny}
  S.~Eidelman and F.~Jegerlehner,
  Z.\ Phys.\ C {\bf 67} (1995) 585
  [hep-ph/9502298].

\bibitem{Groote:2003kg}
  S.~Groote, J.~G.~K\"orner and J.~Maul,
  hep-ph/0309226.

\bibitem{Davier:2010nc}
  M.~Davier, A.~Hoecker, B.~Malaescu and Z.~Zhang,
  Eur.\ Phys.\ J.\ C {\bf 71} (2011) 1515
   Erratum: [Eur.\ Phys.\ J.\ C {\bf 72} (2012) 1874]
  [arXiv:1010.4180 [hep-ph]].

\bibitem{deRafael:1993za}
  E.~de Rafael,
  Phys.\ Lett.\ B {\bf 322} (1994) 239
  [hep-ph/9311316].

\bibitem{Blum:2002ii}
  T.~Blum,
  Phys.\ Rev.\ Lett.\  {\bf 91} (2003) 052001
  [hep-lat/0212018].

\bibitem{DellaMorte:2011aa}
  M.~Della Morte, B.~J\"ager, A.~J\"uttner and H.~Wittig,
  JHEP {\bf 1203} (2012) 055
  [arXiv:1112.2894 [hep-lat]].

\bibitem{Aubin:2012me}
  C.~Aubin, T.~Blum, M.~Golterman and S.~Peris,
  Phys.\ Rev.\ D {\bf 86} (2012) 054509
  [arXiv:1205.3695 [hep-lat]].

\bibitem{Golterman:2014ksa}
  M.~Golterman, K.~Maltman and S.~Peris,
  Phys.\ Rev.\ D {\bf 90} (2014) no.7,  074508
  [arXiv:1405.2389 [hep-lat]].

\bibitem{Bodenstein:2011qy}
  S.~Bodenstein, C.~A.~Dominguez and K.~Schilcher,
  Phys.\ Rev.\ D {\bf 85} (2012) 014029
   Erratum: [Phys.\ Rev.\ D {\bf 87} (2013) no.7,  079902]
  [arXiv:1106.0427 [hep-ph]].

\bibitem{Dominguez:2017omw}
  C.~A.~Dominguez, K.~Schilcher and H.~Spiesberger,
  arXiv:1704.02843 [hep-ph].

\bibitem{Aubin:2006xv}
  C.~Aubin and T.~Blum,
  Phys.\ Rev.\ D {\bf 75} (2007) 114502
  [hep-lat/0608011].

\bibitem{Boyle:2011hu}
  P.~Boyle, L.~Del Debbio, E.~Kerrane and J.~Zanotti,
  Phys.\ Rev.\ D {\bf 85} (2012) 074504
  [arXiv:1107.1497 [hep-lat]].

\bibitem{Feng:2011zk}
  X.~Feng, K.~Jansen, M.~Petschlies and D.~B.~Renner,
  Phys.\ Rev.\ Lett.\  {\bf 107} (2011) 081802
  [arXiv:1103.4818 [hep-lat]].

\bibitem{Francis:2013qna}
  A.~Francis, B.~J\"ager, H.~B.~Meyer and H.~Wittig,
  Phys.\ Rev.\ D {\bf 88} (2013) 054502
  [arXiv:1306.2532 [hep-lat]].

\bibitem{Burger:2013jya}
  F.~Burger {\it et al.} [ETM Collaboration],
  JHEP {\bf 1402} (2014) 099
  doi:10.1007/JHEP02(2014)099
  [arXiv:1308.4327 [hep-lat]].

\bibitem{Chakraborty:2014mwa}
  B.~Chakraborty {\it et al.} [HPQCD Collaboration],
  Phys.\ Rev.\ D {\bf 89} (2014) no.11,  114501
  [arXiv:1403.1778 [hep-lat]].

\bibitem{Chakraborty:2016mwy}
  B.~Chakraborty, C.~T.~H.~Davies, P.~G.~de Oliviera, J.~Koponen 
  and G.~P.~Lepage,
  arXiv:1601.03071 [hep-lat].

\bibitem{Borsanyi:2016lpl}
  S.~Borsanyi {\it et al.},
  arXiv:1612.02364 [hep-lat].

\bibitem{DellaMorte:2017dyu}
  M.~Della Morte {\it et al.},
  arXiv:1705.01775 [hep-lat].

\bibitem{deRafael:2014gxa}
  E.~de Rafael,
  Phys.\ Lett.\ B {\bf 736} (2014) 522
  [arXiv:1406.4671 [hep-lat]].

\bibitem{deRafael:2017gay}
  E.~de Rafael,
  arXiv:1702.06783 [hep-ph].

\bibitem{Benayoun:2016krn}
  M.~Benayoun, P.~David, L.~DelBuono and F.~Jegerlehner,
  arXiv:1605.04474 [hep-ph].

\bibitem{Jegerlehner:2009ry}
  F.~Jegerlehner and A.~Nyffeler,
  Phys.\ Rept.\  {\bf 477} (2009) 1
  [arXiv:0902.3360 [hep-ph]].

\bibitem{Baikov:2008jh}
  P.~A.~Baikov, K.~G.~Chetyrkin and J.~H.~K\"uhn,
  Phys.\ Rev.\ Lett.\  {\bf 101} (2008) 012002
  [arXiv:0801.1821 [hep-ph]].

\bibitem{Penarrocha:2001ig}
  J.~Penarrocha and K.~Schilcher,
  Phys.\ Lett.\ B {\bf 515} (2001) 291
  [hep-ph/0105222].

\bibitem{Beneke:2008ad}
  M.~Beneke and M.~Jamin,
  JHEP {\bf 0809} (2008) 044
  [arXiv:0806.3156 [hep-ph]].

\bibitem{LeDiberder:1992jjr}
  F.~Le Diberder and A.~Pich,
  Phys.\ Lett.\ B {\bf 286} (1992) 147.

\bibitem{Pivovarov:1991rh}
  A.~A.~Pivovarov,
  Z.\ Phys.\ C {\bf 53} (1992) 461
   [Sov.\ J.\ Nucl.\ Phys.\  {\bf 54} (1991) 676]
   [Yad.\ Fiz.\  {\bf 54} (1991) 1114]
  [hep-ph/0302003].

\bibitem{Dominguez:2014fua}
  C.~A.~Dominguez, L.~A.~Hernandez, K.~Schilcher and H.~Spiesberger,
  JHEP {\bf 1503} (2015) 053
  [arXiv:1410.3779 [hep-ph]].

\bibitem{Pich:2016bdg}
  A.~Pich and A.~Rodríguez-Sánchez,
  Phys.\ Rev.\ D {\bf 94} (2016) no.3,  034027
  [arXiv:1605.06830 [hep-ph]].

\bibitem{Bodenstein:2011hm}
  S.~Bodenstein, C.~A.~Dominguez, S.~I.~Eidelman, H.~Spiesberger 
  and K.~Schilcher,
  JHEP {\bf 1201} (2012) 039
  [arXiv:1110.2026 [hep-ph]].

\bibitem{Bodenstein:2013flq}
  S.~Bodenstein, C.~A.~Dominguez, K.~Schilcher and H.~Spiesberger,
  Phys.\ Rev.\ D {\bf 88} (2013) no.1,  014005
  [arXiv:1302.1735 [hep-ph]].

\bibitem{Chetyrkin:2000yt}
  K.~G.~Chetyrkin, J.~H.~Kuhn and M.~Steinhauser,
  Comput.\ Phys.\ Commun.\  {\bf 133} (2000) 43
  [hep-ph/0004189].

\bibitem{Anashin:2015woa}
  V.~V.~Anashin {\it et al.},
  Phys.\ Lett.\ B {\bf 753} (2016) 533
  [arXiv:1510.02667 [hep-ex]].

\bibitem{Anashin:2016hmv}
  V.~V.~Anashin {\it et al.},
  arXiv:1610.02827 [hep-ex].

\bibitem{Boyle:2017gzv}
  P.~Boyle, V.~G\"ulpers, J.~Harrison, A.~J\"uttner, C.~Lehner,
  A.~Portelli and C.~T.~Sachrajda,
  arXiv:1706.05293 [hep-lat].

\bibitem{Giusti:2017jof}
  D.~Giusti, V.~Lubicz, G.~Martinelli, F.~Sanfilippo and S.~Simula,
  arXiv:1707.03019 [hep-lat].


\end{thebibliography}
\end{document}